\documentclass[12pt]{article}
\usepackage{amssymb}
\usepackage{amsmath}
\usepackage{graphicx,color}

\usepackage{graphics}
\usepackage{color}
\usepackage{amssymb,amsmath}
\usepackage{slashed}
\usepackage{epstopdf}
\usepackage{booktabs}
\usepackage{mathrsfs}
\usepackage[export]{adjustbox}
\usepackage{epsfig}
\usepackage{graphicx,color}

\newcommand{\bea}{\begin{eqnarray}}
\newcommand{\eea}{\end{eqnarray}}

\begin{document}
%
\begin{titlepage}
\begin{flushright}
YGHP-17-10
\end{flushright}
\vspace*{10mm}
\begin{center}
\baselineskip 25pt 
{\Large\bf
SU(5)$\times$U(1)$_X$ grand unification with minimal seesaw \\ 
and $Z^\prime$-portal dark matter 
}
\end{center}
\vspace{5mm}
\begin{center}
{\large
Nobuchika Okada$^{~a}$, 
Satomi Okada$^{~b}$
and 
Digesh Raut$^{~a}$
}

\vspace{.5cm}

{\baselineskip 20pt \it
$^a$Department of Physics and Astronomy, University of Alabama, Tuscaloosa, AL35487, USA\\
$^b$Graduate School of Science and Engineering, Yamagata University,\\
Yamagata 990-8560, Japan\\
} 

\end{center}
\vspace{0.5cm}
\begin{abstract}

We propose a grand unified SU(5)$\times$U(1)$_X$ model, 
  where the standard SU(5) grand unified theory is supplemented 
  by minimal seesaw and a right-handed neutrino dark matter 
  with an introduction of a global $Z_2$-parity. 
In the presence of three right-handed neutrinos (RHNs), the model is free from all gauge and mixed-gravitational anomalies. 
The SU(5) symmetry is broken into the Standard Model (SM) gauge group at $M_{\rm GUT} \simeq 4 \times 10^{16}$ GeV 
  in the standard manner, 
  while the U(1)$_X$ symmetry breaking occurs at the TeV scale, 
  which generates the TeV-scale mass of the U(1)$_X$ gauge boson ($Z^\prime$ boson) and the three Majorana RHNs.  
A unique $Z_2$-odd RHN is stable and serves as the dark matter (DM) in the present Universe,   
  while the remaining two RHNs work to generate the SM neutrino masses through the minimal seesaw.  
We investigate the $Z^\prime$-portal RHN DM scenario in this model context. 
We find that the constraints from the DM relic abundance, 
  and the $Z^\prime$ boson search at the Large Hadron Collider (LHC), and the perturbativity bound on the U(1)$_X$ gauge coupling
  are complementary to narrow down the allowed parameter region in the range of  
  $3.0 \leq m_{Z^\prime}[{\rm TeV}] \leq 9.2$ for the $Z^\prime$ boson mass. 
The allowed region for $m_{Z^\prime} \leq 5$ TeV will be fully covered by the future LHC experiments.  
We also briefly discuss the successful implementation of Baryogenesis and cosmological inflation scenarios 
  in the present model.

\end{abstract}
\end{titlepage}


Despite its great success, the Standard Model (SM) suffers from several problems. 
The neutrino mass matrix and a candidate of dark matter (DM) are two major missing pieces 
  of the SM, and they must be supplemented by a framework beyond the SM. 
The minimal U(1)$_{B-L}$ model \cite{MBL} is a very simple extension of the SM, 
  where the anomaly-free global $B-L$ (baryon number minus lepton number) symmetry in the SM is gauged 
  and only the three right-handed neutrinos (RHNs) and the U(1)$_{B-L}$ Higgs field 
  in addition to the SM particle content. 
In the presence of the RHNs, the seesaw mechanism \cite{seesaw} is automatically implemented.  
Associated with the spontaneous U(1)$_{B-L}$ symmetry breaking by a vacuum expectation value (VEV) 
  of the U(1)$_{B-L}$ Higgs field, the U(1)$_{B-L}$ gauge boson ($Z^\prime$ boson) and   
  the three Majorana RHNs acquire their masses. 
The SM neutrino mass matrix is generated through the seesaw mechanism after the electroweak symmetry breaking.

Among various possibilities of introducing a DM candidate into the minimal U(1)$_{B-L}$ model, 
  a way proposed in Ref.~\cite{OS} would be the simplest, where instead of extending the particle content, 
  a $Z_2$-parity is introduced and a unique $Z_2$-odd RHN plays the role of DM. 
The remaining two RHNs work to generate the SM neutrino mass matrix through the seesaw mechanism. 
Therefore, in this framework, the three RHNs are categorized into one $Z_2$-odd RHN DM 
  and two $Z_2$-even RHNs for the so-called Minimal Seesaw \cite{MinSeesaw}, 
  which is the minimal setup to reproduce the neutrino oscillation data 
  with a prediction of one massless eigenstate.    
In this way, the two missing pieces of the SM are supplemented 
  with no extension of the particle content of the minimal U(1)$_{B-L}$ model.

The RHN DM communicates with the SM particles in two ways: 
One is through exchanges of Higgs bosons in their mass basis (Higgs-portal RHN DM), 
  where two physical Higgs bosons are realized as linear combinations of 
  the U(1)$_{B-L}$ and the SM Higgs bosons. 
The other is through $Z^\prime$ boson exchange ($Z^\prime$-portal RHN DM). 
Phenomenology of the RHN DM has been extensively studied \cite{OS, RHNDM1, OO1}, 
  in particular, a complementarity between the RHN DM physics and the LHC physics 
  for the $Z^\prime$-portal RHN DM scenario has been pointed out in Ref.~\cite{OO1}.

The minimal U(1)$_{B-L}$ model is easily generalized to the minimal U(1)$_X$ model \cite{MinU1X}, 
  whose particle content is the same as the one of the minimal U(1)$_{B-L}$ model. 
The generalization appears in the U(1)$_X$ charge assignment for the SM fields: 
  the U(1)$_X$ charge of an SM field ($f$) is defined as $Q_X= Y_f \, x_H + Q^f_{B-L}$, 
  where $Y_f$ and $Q^f_{B-L}$ are the hypercharge and the $B-L$ charge of the field, 
  and $x_H$ is a new real parameter (see Table~\ref{table1}). 
With the charge assignment, the minimal U(1)$_X$ model is free from all the gauge and mixed-gravitational anomalies 
  (see, for example, Ref.~\cite{OOT} for detailed calculations of the anomaly coefficients).
The minimal U(1)$_{B-L}$ model is defined as the limit of $x_H \to 0$. 
The RHN DM is introduced in exactly the same way for the U(1)$_{B-L}$ model. 
In Ref.~\cite{OO2}, the minimal U(1)$_X$ model with the $Z^\prime$-portal RHN DM 
   has been extensively studied. 
In the analysis, only four free parameters are involved: 
    the U(1)$_X$ gauge coupling ($\alpha_X$), the $Z^\prime$ boson mass ($m_{Z^\prime}$), 
    $x_H$, and the RHN DM mass ($m_{\rm DM}$).  
It has been found in Ref.~\cite{OO2} that $m_{\rm DM} \simeq m_{Z^\prime}/2 $ is required 
   to reproduce the observed DM relic density, and the number of free parameters 
   is effectively reduced to three:  $\alpha_X$, $m_{Z^\prime}$ and $x_H$. 
The cosmological constraint from the DM relic density leads to a lower bound on $\alpha_X$ 
   for fixed values of $m_{Z^\prime}$ and $x_H$. 
On the other hand,  the LHC Run-2 results from the search for a $Z^\prime$ boson resonance 
   provide an upper bound on $\alpha_X$ for fixed values of $m_{Z^\prime}$ and $x_H$. 
Therefore, the DM physics and the LHC Run-2 phenomenology are complementary 
   to narrow down the model parameter space.

In this letter, we propose a grand unified SU(5)$\times$U(1)$_X$ model,\footnote{
A similar model, but 
the unification into the flipped SU(5)$\times$U(1) group has been recently proposed in Ref.~\cite{Chen:2017rpn}. 
}
 into which 
   the minimal U(1)$_X$ model with the RHN DM is embedded. 
The grand unified theory (GUT) has been attracting a lot of attention since its first proposal in Ref.~\cite{GUT}, 
   where all the three gauge interactions in the SM are embedded into the SU(5) gauge group, 
   and all the fermions in the SM are unified into three generations of ${\bf 5^*}$ and ${\bf 10}$-representations 
   under the SU(5).  
The picture is not only mathematically beautiful, but also provides the charge quantization for the SM quarks and leptons. 
It seems natural to regard the GUT as a primary candidate of physics beyond the SM. 
However, if this is the case, we may require a GUT model to incorporate the neutrino mass matrix 
   and a DM candidate. 
The model we propose satisfies this requirement with the RHN DM and two RHNs for the minimal type-I seesaw. 
As in the original proposal in Ref.~\cite{GUT}, the SM gauge groups are embedded into the SU(5) group. 
However, note that the unification of the quarks and leptons into ${\bf 5^*}$ and ${\bf 10}$-representations 
   is possible only if $x_H=-4/5$. 
Therefore, the U(1)$_X$ charge is quantized and $x_H$ is no longer a free parameter.

In the following, we show that our GUT model is phenomenologically viable. 
As we will discuss below, the SU(5) gauge symmetry is broken at $M_{\rm GUT} \simeq 4 \times 10^{16}$ GeV,  
  and the minimal U(1)$_X$ model with the RHN DM ($x_H=-4/5$) is realized as low energy effective theory. 
The U(1)$_X$ symmetry is assume to be broken at the TeV scale. 
We first review this effective minimal U(1)$_X$ model at the TeV scale, 
   and investigate phenomenological constraints on the model parameters. 
In our analysis, we follow Ref.~\cite{OO2} and identify an allowed parameter region, 
   which will be found to be very narrow since $x_H= -4/5$ is no longer a free parameter. 
Next, we discuss that the SM gauge couplings are successfully unified at $M_{\rm GUT}$ 
   with some extra fermions at the TeV scale, which originate 
   one ${\bf 5}+{\bf 5^*}$ and one ${\bf 10}+{\bf 10^*}$ multiplets under the SU(5) gauge group.  
After the SU(5) breaking, a kinetic mixing between the U(1)$_Y$ and U(1)$_X$ gauge bosons 
   is generated through the renormalization group (RG) evolution. 
We also discuss that this mixing is negligibly small and has little effect on our analysis.

\begin{table}[t]
\begin{center}
\begin{tabular}{c|ccc|c|c}
      &  SU(3)$_C$  & SU(2)$_L$ & U(1)$_Y$ & U(1)$_X$  & $ Z_2 $\\ 
\hline
$q^{i}_{L}$ & {\bf 3 }    &  {\bf 2}         & $ 1/6$       & $(1/6) x_{H}   +1/3  $    & $+$\\
$u^{i}_{R}$ & {\bf 3 }    &  {\bf 1}         & $ 2/3$       & $(2/3) x_{H}  +1/3 $   & $+$\\
$d^{i}_{R}$ & {\bf 3 }    &  {\bf 1}         & $-1/3$       & $(-1/3) x_{H} +1/3 $   &$+$\\
\hline
$\ell^{i}_{L}$ & {\bf 1 }    &  {\bf 2}         & $-1/2$       & $(-1/2) x_{H} - 1$  & $+$ \\
$e^{i}_{R}$    & {\bf 1 }    &  {\bf 1}         & $-1$                   & $(-1) x_{H} - 1$   & $+$ \\
\hline
$H$            & {\bf 1 }    &  {\bf 2}         & $- 1/2$       & $(-1/2) x_{H}$  & $+$ \\  
\hline
$N^{j}_{R}$    & {\bf 1 }    &  {\bf 1}         &$0$                    & $- 1$   & $+$ \\
$N_{R}$         & {\bf 1 }    &  {\bf 1}         &$0$                    & $- 1$   & $-$ \\
$\Phi$            & {\bf 1 }       &  {\bf 1}       &$ 0$                  & $ + 2$  & $+$ \\
\hline
\end{tabular}
\end{center}
\caption{
The particle content of the minimal U(1)$_X$ extended SM with $Z_2$-parity. 
In addition to the SM particle content ($i=1,2,3$), the three RHNs 
  ($N_R^j$ ($j=1,2$) and $N_R$) and the U(1)$_X$ Higgs field ($\Phi$) are introduced.   
The unification into SU(5)$\times$U(1)$_X$ is achieved only for $x_H=-4/5$, and 
  $x_H$ is quantized in our model. 
}
\label{table1}
\end{table}

We first define the minimal U(1)$_X$ model by the particle content listed in Table~\ref{table1}. 
The minimal $B-L$ model is reproduced as the limit of $x_H \to 0$. 
The model is free from all the gauge and the gravitational anomalies 
  in the presence of the three RHNs.  
Because of the $Z_2$-parity assignment shown in Table~\ref{table1}, the $N_R$ is a unique (cold) DM candidate. 
Fixing $x_H=-4/5$, we can see the unification of quarks and lepton into SU(5)$\times$U(1)$_X$ multiplets: 
$\overline{F_{5}}^i$ of $({\bf 5^*}, -3/5) \supset (d^{i}_{R})^c \oplus \ell^{i}_{L}$, 
and 
$F_{10}^i$ of $({\bf 10}, 1/5) \supset q^{i}_L \oplus (u^{i}_{R})^c \oplus  (e^{i}_{R})^c$.

The Yukawa sector of the SM is extended to have 
\bea
\mathcal{L}_{Yukawa} &\supset&  - \sum_{i=1}^{3} \sum_{j=1}^{2} Y^{ij}_{D} \overline{\ell^i_{L}} H N_R^j 
          -\frac{1}{2} \sum_{k=1}^{2} Y^k_N \Phi \overline{N_R^{k~C}} N_R^k    
     - \frac{1}{2}  Y_N \Phi \overline{N_R^{~C}} N_R + {\rm H.c.} ,
\label{Lag1} 
\eea
where the first term is the neutrino Dirac Yukawa coupling, and the second and 
   third terms are the Majorana Yukawa couplings. 
Without loss of generality, the Majorana Yukawa couplings are already diagonalized in our basis.  
Note that only the two $Z_2$-even RHNs are involved 
  in the neutrino Dirac Yukawa coupling. 
After the U(1)$_X$ and the electroweak symmetry breakings, 
  the $Z^\prime$ boson mass, the Majorana masses for the RHNs, and the neutrino Dirac masses are generated:
\bea
  && m_{Z^\prime} = g_X \sqrt{4 v_\Phi^2+  \frac{1}{4}x_H^2 v_h^2} \simeq 2 g_X v_\Phi , \nonumber \\
  && m_{N^i}=\frac{Y_N^i}{\sqrt{2}} v_\Phi, \; \; 
        m_{\rm DM}=\frac{Y_N}{\sqrt{2}} v_\Phi,   
  \; \; m_{D}^{ij}=\frac{Y_{D}^{ij}}{\sqrt{2}} v_h,
  \eea   
where $g_X$ is the U(1)$_X$ gauge coupling,  $\langle \Phi \rangle = v_\Phi/\sqrt{2}$, $v_h = 246$ GeV 
   is the SM Higgs VEV, and we have used the LEP constraint \cite{Carena:2004xs} ${v_\Phi}^2 \gg {v_h}^2$.

Now we investigate phenomenological constraints for the minimal U(1)$_X$ model with the RHN DM. 
We follow Ref.~\cite{OO2} for our analysis, where only four free parameters, 
  $\alpha_X=g_X^2/(4 \pi)$, $m_{Z^\prime}$,  $x_H$ and $m_{\rm DM}$. 
However, the grand unification to SU(5)$\times$U(1)$_X$ requires to fix $x_H=-4/5$, 
  and hence the resultant allowed parameter space is more restricted.

Let us first evaluate the DM relic density by integrating the Boltzmann equation given by \cite{KolbTurner}
\bea 
  \frac{dY}{dx}
  = - \frac{x s \langle \sigma v_{\rm rel} \rangle}{H(m_{\rm DM})} \left( Y^2-Y_{EQ}^2 \right), 
\label{Boltmann}
\eea  
where $x=m_{\rm DM}/T$ is the ratio of the DM mass to the temperature of the Universe ($T$), 
  $H(m_{\rm DM})$ is the Hubble parameter at $T=m_{\rm DM}$, 
  $Y$ is the yield (the ratio of the DM number density to the entropy density $s$) of 
  the DM, and $Y_{EQ}$ is the yield of the DM particle in thermal equilibrium. 
The thermal average of the DM annihilation cross section times relative velocity, 
  $\langle \sigma v_{\rm rel} \rangle$,  is calculated from the total cross section 
  of the DM pair annihilation process $NN \to Z^\prime \to f {\bar f}$ 
   ($f$ denotes the SM fermions) given by
\bea 
 \sigma(s)=\frac{\pi}{3}  \alpha_X^2  \frac{\sqrt{s (s-4 m_{\rm DM}^2)}}
  {(s-m_{Z^\prime}^2)^2+m_{Z^\prime}^2 \Gamma_{Z^\prime}^2} 
    F(x_H),  
\label{DMSigma}
\eea 
where 
\bea 
  F(x_H) = 10  \left( x_H+ \frac{4}{5} \right)^2 + \frac{33}{5}, 
\label{F}  
\eea
  and the total decay width of $Z^\prime$ boson is given by 
\bea
\Gamma_{Z'} &=& 
  \frac{\alpha_X}{6} m_{Z^\prime} 
  \left[ F(x_H) + \beta^3 \theta(\beta^2)
 \right], 
\label{width}
\eea
  where $\beta=\sqrt{1-4 m_{\rm DM}^2/m_{Z^\prime}^2}$, $\theta(z)$ is the unit step function, 
  and the masses of all SM fermions are neglected. 
Here, we have assumed $m_{N_R^{1,2}} > m_{\rm DM},$ $m_{Z^\prime}/2$,  for simplicity. 
By solving the Boltzmann equation numerically, 
  the present DM relic density is evaluated by 
\bea 
  \Omega_{DM} h^2 =\frac{m_{\rm DM} \, s_0 \, Y(\infty)} {\rho_c/h^2}, 
\eea 
  where $s_0 = 2890$ cm$^{-3}$ is the entropy density of the present Universe, 
  and $\rho_c/h^2 =1.05 \times 10^{-5}$ GeV/cm$^3$ is the critical density. 
We identify a parameter region to reproduce the observed DM relic density 
  in the range of $0.1183 \leq \Omega_{\chi} h^2 \leq 0.1213$ (68\% confidence level), 
  measured by the Planck satellite experiment~\cite{Ade:2015xua}. 
As shown in Ref.~\cite{OO2}, an enhancement of the DM pair annihilation cross section 
  via the $Z^\prime$ boson resonance in the $s$-channel necessary 
  to reproduce the observed DM density.   
Hence, we always find $m_{\rm DM} \simeq m_{Z^\prime}/2$, and thus 
  our results are effectively described by only two free parameters: $\alpha_{X}$ and $m_{Z^\prime}$.

\begin{figure}[t]
\begin{center}
\includegraphics[width=0.6\textwidth,angle=0,scale=1.2]{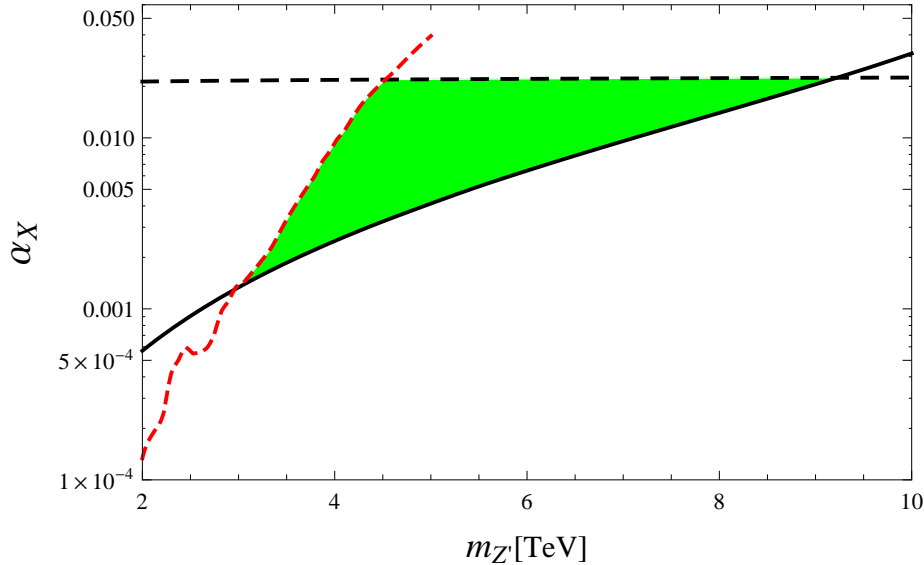} 
\end{center}
\caption{
Allowed parameter region (green shaded) for the $Z^\prime$-portal RHN DM scenario 
  in the context of  our SU(5)$\times$U(1)$_X$ model ($x_H=-4/5$). 
The (black) solid line denotes the lower bound on $\alpha_X$ as a function of $m_{Z^\prime}$ 
  to reproduce the observed DM relic abundance. 
The diagonal dashed line (in red) shows the upper bound on $\alpha_X$ obtained 
  from the search results for a $Z^\prime$ boson resonance at the LHC, 
  which is applicable to $m_{Z^\prime} \leq 5.0$ TeV.   
The perturbativity bound (see the discussion around Eq.~(\ref{PTB})) is depicted by the horizontal dashed line.    
Combining all three constraints, we obtain the $Z^\prime$ boson mass bound 
  in the range of $3.0 \leq m_{Z^\prime}[{\rm TeV}] \leq 9.2$. 
}
\label{fig:1}
\end{figure}

Next we consider the current LHC constraints from the search 
  for a narrow resonance with dilepton final states at the LHC Run-2. 
In the analysis by the ATLAS and CMS collaborations, 
  the so-called sequential SM $Z^\prime$ ($Z^\prime_{SSM}$) has been studied as a reference model. 
We interpret the current LHC constraints on the $Z^\prime_{SSM}$ boson 
  into the $Z^\prime$ boson in our U(1)$_X$ model to identify an allowed parameter region.  
Although our analysis follows that in Ref.~\cite{OO2}, 
  we will update the LHC constraints presented in Ref.~\cite{OO2} 
  by employing the latest ATLAS results in 2017 \cite{ATLAS:2017}.

The cross section for the process $pp \to Z^\prime +X \to \ell^{+} \ell^{-} +X$  is given by
\begin{eqnarray}
 \sigma
 &=& 2 \, \sum_{q, {\bar q}}
 \int d M_{\ell \ell} 
 \int^1_ \frac{M_{\ell \ell}^2}{s} dx
 \frac{2 M_{\ell \ell}}{x s}  
 f_q(x, Q^2) f_{\bar q} \left( \frac{M_{\ell \ell}^2}{x s}, Q^2
 \right)    
 {\hat \sigma} (q \bar{q} \to Z^\prime \to  \ell^+ \ell^-) ,
\label{X_LHC}
\end{eqnarray}
where $M_{\ell \ell}$ is the invariant mass of a final state dilepton,  
  $f_q$ is the parton distribution function for a parton (quark) ``$q$'', 
  and $\sqrt{s} =13$ TeV is the center-of-mass energy of the LHC Run-2.
In our numerical analysis, we employ CTEQ6L~\cite{CTEQ} 
  for the parton distribution functions with the factorization scale $Q= m_{Z^\prime}$. 
The cross section for the colliding partons is given by 
\bea 
{\hat \sigma}(q \bar{q} \to Z^\prime \to  \ell^+ \ell^-) = \frac{\pi}{1296} \, \alpha_X^2 \, 
\frac{M_{\ell \ell}^2}{(M_{\ell \ell}^2-m_{Z^\prime}^2)^2 + m_{Z^\prime}^2 \Gamma_{Z^\prime}^2} 
F_{q \ell}(x_H),  
\label{CrossLHC2}
\eea
where the function $F_{q \ell}(x_H)$ is given by 
\bea
   &&F_{u \ell}(x_H) =  (8 + 20 x_H + 17 x_H^2)  (8 + 12 x_H + 5 x_H^2),   \nonumber \\
   &&F_{d \ell}(x_H) =  (8 - 4 x_H + 5 x_H^2) (8 + 12 x_H + 5 x_H^2) 
\label{Fql}
\eea
for $q$ being the up-type ($u$) and down-type ($d$) quarks, respectively.  
By integrating the differential cross section over a range of $M_{\ell \ell}$ set by the ATLAS analysis, 
  we obtain the cross section to be compared with the upper bounds 
  obtained by the ATLAS collaboration. 
Only two free parameters, $\alpha_{X}$ and $m_{Z^\prime}$, are involved in our analysis. 

\begin{figure}[h]
\begin{center}
\includegraphics[width=0.6\textwidth,angle=0,scale=1.2]{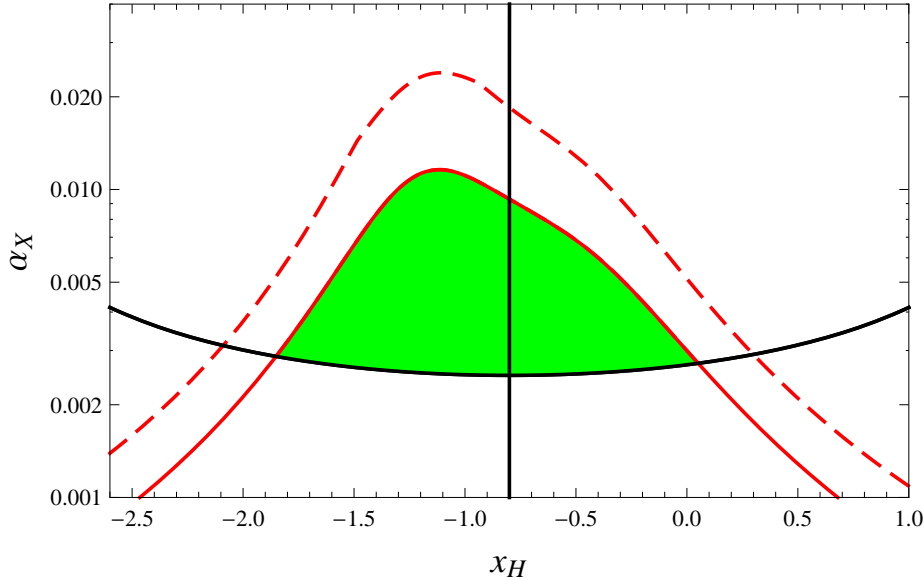} 
\end{center}
\caption{
Allowed parameter region (green shaded)  as a function of $x_H$, for $m_{Z^\prime}=4$ TeV 
  in the minimal U(1)$_X$ model with the $Z^\prime$-portal RHN DM. 
For the grand unified model, $x_H=-4/5$ is indicated by a vertical solid line.     
The (black) solid line shows the cosmological lower bound on $\alpha_X$ as a function of $x_H$. 
The dashed line (in red) shows the upper bound on $\alpha_X$ 
   obtained in Ref.~\cite{OO2} by employing the LHC data with a 13/fb luminosity. 
Our update of the results by employing the ATLAS results with 36/fb 
   is shown by the (red) solid line. 
}
\label{fig:2}
\end{figure}

In Figure~\ref{fig:1}, we show our combined results from the DM relic abundance and the search results 
  for a $Z^\prime$ boson resonance at the LHC. 
We can see these two constraints are complementary 
  to narrow down the allowed parameter region. 
In Figure~\ref{fig:1}, we also show the perturbativity bound (see the discussion around Eq.~(\ref{PTB})), 
  which provides the upper bound on the U(1)$_X$ gauge coupling. 
Combining all three constraints, we have obtained the $Z^\prime$ boson mass bound 
  in the range of $3.0 \leq m_{Z^\prime}[{\rm TeV}] \leq 9.2$.

Although in our grand unified model $x_H=-4/5$ is not a free parameter, we present the combined results as a function of $x_H$ in Figure~\ref{fig:2} for $m_Z^\prime=4$ TeV, in order to show that $x_H= -4/5$ has an interesting phenomenological implication.  
The (black) convex-downward solid line shows the cosmological lower bound on $\alpha_X$ as a function of $x_H$. 
The (red) convex-upward dashed line shows the upper bound on $\alpha_X$ 
  presented in Ref.~\cite{OO2}, where the results are obtained from the LHC 2016 data with a 13/fb luminosity. 
We have updated the results by employing the latest ATLAS results with 36/fb \cite{ATLAS:2017}, 
  and our result is shown by the (red) convex-upward solid line. 
The (green) shaded region is the final result for the allowed parameter space 
   after combining the cosmological and the LHC constraints when $x_H$ is a free parameter. 
Interestingly, the plot indicates that $x_H=-4/5$ required by the grand unification into SU(5)$\times$U(1)$_X$ 
   is almost the best value within the allowed region.

Let us now consider the gauge coupling unification. 
To realize the grand unification picture, the SM gauge coupling must be unified 
  at a high energy scale. 
It has been shown in Ref.~\cite{GCU} that the SM gauge couplings are successfully unified 
  around $M_{\rm GUT} \simeq 4 \times 10^{16}$ GeV 
  in the presence of two pairs of vector-like quarks,
  $D_L+D_R$ and $Q_L+Q_R$, with their mass of ${\cal O}$(1 TeV)
  in the representations of $({\bf 3}, {\bf 1}, 1/3)$ and $({\bf 3},{\bf 2}, 1/6)$, 
  respectively, under the SM gauge group of SU(3)$_C\times$SU(2)$_L\times$U(1)$_Y$. 
Here, we adopt this simple case to our scenario with a common mass ($M$) 
  for $D_L+D_R$ and $Q_L+Q_R$. 
We simply take $M=m_{Z^\prime}/2$ not to change the $Z^\prime$ boson decay width 
   that we have used in our DM analysis. 
The unification scale leads to proton lifetime to be $\tau_p \simeq 10^{38}$ years, 
   which is consistent with the current experimental lower bound obtained by the Super-Kamiokande \cite{Proton_life}, 
  $\tau_p(p \to \pi^0 e^+) \gtrsim 10^{34}$ years.

In the grand unified SU(5)$\times$U(1)$_X$ picture, 
  $D_L+D_R^{~C}$ and $Q_L+Q_R^{~C}$ 
  are unified into $F_5+\overline{F_5}=({\bf 5}, 3/5)+({\bf 5^*}, -3/5)$ 
  and  $F_{10}+\overline{F_{10}}=({\bf 10}, 1/5)+({\bf 10^*}, -1/5)$, respectively. 
The way to realize the mass splittings among the components in the SU(5) multiplets 
  and leave only $D_L+D_R$ and $Q_L+Q_R$ light 
  is analogous to the triplet-doublet mass splitting for the ${\bf 5}$-plet Higgs field 
  in the usual SU(5) GUT model.  
We introduce the Yukawa couplings and the mass terms such as  
\bea
  {\cal L} _Y = \overline{F_{5}} (Y_5 \Sigma  -M_5 )F_{5}  +
  {\rm tr} \left[\overline{F_{10}} (Y_{10} \Sigma  -M_{10})F_{10} \right],  
\eea
where $Y_{5, 10}$ are Yukawa coupling constants, 
  $M_{5,10}$ are masses for the vector-like fermions, 
  $\Sigma$ is a U(1)$_X$ charge-neutral SU(5) adjoint Higgs field, 
  whose VEV of $\langle \Sigma \rangle = v_{\rm GUT} \, {\rm diag}(1,1,1,-3/2, -3/2)$ 
  breaks the SU(5) gauge group into the SM ones, 
  and we have used antisymmetric $5 \times 5$ matrices to express $F_{10}$ and $\overline{F_{10}}$. 
We tune the Yukawa coupling to realize $Y_5 v_{\rm GUT} -M_5={\cal O}$(1 TeV) for $F_5+\overline{F_5}$, 
  so that the vector-like SU(3)$_C$ color triplets ($D_L+D_R$) become light, while the vector-like SU(2)$_L$ doublets 
  in the multiplets are heavy with mass of $(5/2) M_5$. 
By tuning $Y_{10}$ to realize $Y_{10} v_{\rm GUT} + 4 M_{10} = {\cal O}$(1 TeV) for $F_{10}$ and $\overline{F_{10}}$, 
  we obtain $Q_L+Q_R$ light, while the rest in the {\bf 10}-plets are heavy with mass of $ 5 M_{10}$.
Taking suitable values for $Y_{5, 10}={\cal O}$(1), we can obtain the common TeV scale mass for $D_L+D_R$ and $Q_L+Q_R$, 
  while the other components have GUT-scale masses.

Once the SU(5) symmetry is broken to the SM gauge groups at $M_{\rm GUT}$, 
   a kinetic mixing between the U(1)$_Y$ and the U(1)$_X$ gauge bosons is generated at low energies 
   through the RG evolutions. 
Following standard techniques in Ref.~\cite{delAguila:1988jz}, 
   we generally set a basis where the gauge boson kinetic terms are diagonalized 
   and a covariant derivative of a field is defined as 
\bea
  D_\mu = \partial_\mu - 
  \left(Y \; Q_X \right)
 \begin{pmatrix} 
     g_Y  & g_{\rm mix} \\
     0      &  g_X  \\
   \end{pmatrix}
 \begin{pmatrix} 
     B_\mu  \\
     Z^\prime_\mu \\
   \end{pmatrix}   .
\eea
Here, $Y$ and $Q_X$ are U(1)$_Y$ and U(1)$_X$ charges of the field, respectively, 
  $B_\mu$ is the SM U(1)$_Y$ gauge field, and $g_Y$ is the U(1)$_Y$ gauge coupling. 
Originating from the gauge kinetic mixing,  
  a new parameter, namely, ``mixed gauge coupling'' $g_{\rm mix}$ is introduced. 
In this basis, the RG evolution of  the SM U(1)$_Y$ gauge coupling remains the same as the SM one 
  at the one-loop level, while $g_X$ and $g_{\rm mix}$ evolve according to their coupled RG equations.  
At the one-loop level, the coupled RG equations for $\mu > {\cal O}$(TeV) are given by 
\bea 
  \mu \frac{d g_X}{d \mu} = \frac{ \beta_{g_X}}{16 \pi^2} , \; \; \;
  \mu \frac{d g_{\rm mix}}{d \mu} =\frac{\beta_{g_{\rm mix}}}{16 \pi^2} ,
\eea   
where 
\bea
  \beta_{g_X} &=& \frac{1}{6} g_X 
  \left[ 
  \left( 80 + 64 x_H + 45 x_H^2 \right) g_X^2 
  +   2 \left( 32 + 45 x_H \right) g_X g_{\rm mix} + 45 g_{\rm mix}^2  \right], \nonumber \\
\beta_{g_{\rm mix}} 
& = & \frac{5}{3}g_Y^2 \left[  \left( \frac{32}{5} + 9 x_H \right)  g_X   + 9 g_{\rm mix}  \right]  \nonumber\\
 &+&  \frac{1}{6} g_{\rm mix}  \left[ 
     (80 + 64 x_H + 45 x_H^2) g_X^2  +  2 (32 + 45 x_H) g_X g_{\rm mix}  + 45 g_{\rm mix}^2  \right].  
\eea
Here we have taken into account all particle contributions to the beta functions at the TeV scale. 
Numerically solving the RG equations with $g_{\rm mix}=0$ and various values of $g_X$ at $\mu=M_{\rm GUT}$,  
  we have found that $g_{\rm mix}/g_X \simeq 0.034$ at the TeV scale 
 for any input values of $g_X$ at $M_{\rm GUT}$. 
Therefore, we can safely neglect effects of $g_{\rm mix}$ in our analysis and set $g_{\rm mix}=0$ 
 as a good approximation. 

Neglecting $g_{\rm mix}$ in the RG equations, we find the following analytic solution for the U(1)$_X$ gauge coupling: 
\bea
  \alpha_X(m_{Z^\prime}) = \frac{\alpha_X(M_{\rm GUT})}{1+ \alpha_X(M_{\rm GUT}) \frac{b_X}{2 \pi}  \ln\left[\frac{M_{\rm GUT}}{m_{Z^\prime}} \right]}, 
\label{PTB}
\eea
where $b_X= \left( 80 + 64 x_H + 45 x_H^2 \right)/6 = 48/5$ for $x_H=-4/5$.  
Since the running U(1)$_X$ gauge coupling $\alpha_X(\mu)$ is asymptotically non-free, 
  we now impose the ``perturbativity bound'' that $\alpha_X(M_{\rm GUT})$ must be in the perturbative regime. Adopting a condition of $\alpha_X(M_{\rm GUT}) \leq 4 \pi$, we find $\alpha_X(m_{Z^\prime}) \leq 0.022$ for $m_{Z^\prime} \leq 10$ TeV. 
In Figure~\ref{fig:1}, we see that this perturbativity bound is more severe than the LHC bound 
  for $m_{Z^\prime} \gtrsim 4.5$ TeV.

Finally, our grand unified SU(5)$\times$U(1)$_X$ model can also account for the origin of the Baryon asymmetry in the Universe 
  through leptogenesis \cite{FY} with two $Z_2$-even RHNs 
  if they are almost degenerate (resonant leptogenesis \cite{ResLG}).  
Introducing non-minimal gravitational couplings, the U(1)$_X$ Higgs field plays the role of inflaton. 
We can achieve the successful cosmological inflation scenario with a suitable choice of the non-minimal gravitational coupling constant.   
See, for example, Ref.~\cite{BL-Inflation}.

\section*{Acknowledgments}
This work of N.O. is supported in part by the U.S. Department of Energy (DE-SC0012447).



\end{document}